\documentclass[conference,letterpaper]{IEEEtran}
\usepackage{amsfonts,amssymb,amsmath,amsthm,mathtools,cancel}
\usepackage{color}
\usepackage{graphicx}
\usepackage{subcaption}
\usepackage{anyfontsize}
\RequirePackage{fix-cm}

\newcommand{\mb}{\mathbf}
\newcommand{\mbb}{\mathbb}

\newcommand{\ora}{\overrightarrow}
\newcommand{\rmn}[1]{\uppercase\expandafter{\romannumeral #1\relax}}
\newcommand{\R}{\mathbb{R}}
\newcommand{\N}{\mathbb{N}}
\newcommand{\C}{\mathbb{C}}
\newcommand{\tr}{\text{\hspace{-1pt}\raisebox{.5pt}{\scalebox{.85}{$\top$}}}} 
\newenvironment{spmatrix}{
    \left(
    \begin{smallmatrix}
    }
    {
    \end{smallmatrix}\right)}
\theoremstyle{plain}

\theoremstyle{remark}

\begin{document}
%
\title{Quantum Illumination Advantage for Classification Among an Arbitrary Library of Targets}
%
%
%


\author{\IEEEauthorblockN{Ali Cox}
		\IEEEauthorblockA{College of Optical Sciences\\ Univ. of Arizona}
  \and
  		\IEEEauthorblockN{Quntao Zhuang}
		\IEEEauthorblockA{Electrical and Computer Engineering\\ Univ. of Southern California}
  \and
  		\IEEEauthorblockN{Jeffrey H. Shapiro}
		\IEEEauthorblockA{Res. Lab. of Electronics\\ Mass. Inst. of Technology}

  \and
		\IEEEauthorblockN{Saikat Guha}
		\IEEEauthorblockA{College of Optical Sciences\\ Univ. of Arizona}
	}

%
%

\markboth{Journal of \LaTeX\ Class Files,~Vol.~14, No.~8, August~2015}%
{Shell \MakeLowercase{\textit{et al.}}: Bare Demo of IEEEtran.cls for IEEE Journals}
%



\maketitle

\begin{abstract}
    Quantum illumination (QI) is the task of querying a scene using a transmitter probe whose quantum state is entangled with a reference beam retained in ideal storage, followed by optimally detecting the target-returned light together with the stored reference, to make decisions on characteristics of targets at stand-off range, at precision that exceeds what is achievable with a classical transmitter of the same brightness and otherwise identical conditions. Using tools from perturbation theory, we show that in the limit of low transmitter brightness, high loss, and high thermal background, there is a factor of four improvement in the Chernoff exponent of the error probability in discriminating any number of apriori-known reflective targets when using a Gaussian-state entangled QI probe, over using classical coherent-state illumination (CI). While this advantage was known for detecting the presence or absence of a target, it had not been proven for the generalized task of discriminating between arbitrary target libraries. In proving our result, we derive simple general analytic expressions for the lowest-order asymptotic expansions of the quantum Chernoff exponents for QI and CI in terms of the signal brightness, loss, thermal noise, and the modal expansion coefficients of the target-reflected light's radiant exitance profiles when separated by a spatial mode sorter after entering the entrance pupil of the receiver's aperture.
\end{abstract}


%
\IEEEpeerreviewmaketitle

\section{Introduction}
%
%
%
%

\IEEEPARstart{T}{here} has been considerable interest in the performance advantage (measured by the quantum Chernoff exponent) under a signal mean photon number constraint  of illumination using a signal-idler pair excited in an entangled Gaussian state over illumination using a signal excited in a classical coherent state. The literature investigating this advantage of Gaussian quantum illumination (GQI) over classical coherent illumination (CI) started in 2008 when it was shown \cite{Tan08} that using GQI over CI results in a significant performance boost, for the simplest form of stand-off discrimination: determining the presence or absence of a target expected to be in a known region of space with a high amount of thermal background noise and high round-trip loss. In the extreme limit of high loss, high noise and weak signal brightness, the quantum Chernoff exponent under GQI was shown to be exactly 4 times greater that under CI. Structured receiver proposals followed \cite{Guh09, Bar15} that strived to achieve this full factor of 4 improvement over the best possible receiver for coherent state illumination, but the proposed designs only promised to achieve half of this advantage.

It wasn't until 2017, when a receiver was proposed \cite{Zhu17} capable of fully closing the gap between CI and GQI performance. This receiver requires using a complicated non-Gaussian operation to convert target-presence information encoded in the correlation of the return-idler radiation fields into a displacement of an auxiliary sum-frequency mode. A more simple method based on Heterodyne detection was only recently proposed \cite{Shi22} to perform the same correlation-to-displacement conversion to achieve the 4-fold advantage of GQI.

With the exception of a numerical exploration in a simple one-dimensional setting of discriminating between one or two point targets \cite{Guh10}, and a quantum ranging study \cite{Zhu21}, where the correct target range is encoded in only one of many possible target return signal modes, the existing literature on the topic of quantum illumination has exclusively focused on the advantage of GQI over CI for determining the presence or absence of a target. In this paper, we take on the most general task of discriminating among a known set of targets of arbitrary shapes and multiplicity. We show analytically that in the extreme limit of high thermal background noise, high loss and low signal brightness, the four-fold gain in the quantum Chernoff exponent afforded by GQI over CI extends into this generalized setting of multiple arbitrary targets.

In deriving this result, we provide a necessary scaling relation among the loss, thermal noise and signal brightness in the regime where the advantage occurs, and derive asymptotic approximations of the quantum Chernoff exponents for GQI and CI in terms of the decomposition coefficients of the target return light into the fixed spatial mode basis on which an optimal receiver operates. Such a proof is normally complicated by the highly multi-modal representation of the state of the target return light  in order to accurately capture the shape-information that uniquely characterizes each target in the library. Although the target return light is excited in a Gaussian state (in the case of CI and GQI alike), for which the quantum Chernoff bound is known to be computable \cite{Pir08}, an analytical evaluation remains out of reach due to the intractability of analytic symplectic diagonalization (required by the Gassian quantum Chernoff exponent formula) of a large covariance matrix for multi-modal states. To overcome this challenge, we take advantage of the high-loss, high noise and weak signal regime to approximate the return state as a fixed zeroth order state $\hat{\rho}^{(0)}$ plus a small traceless target-dependent perturbation $\hat{\nu}^{(i)}$ proportional to the product of loss and signal-to-noise ratio. We then apply perturbation theoretic methods from \cite{Grace22} to obtain an asymptotic expansion of the quantum Chernoff bounds in the regime of interest.
\section{Problem Setup}
\begin{figure}[t]
\centering
\includegraphics[width=\linewidth]{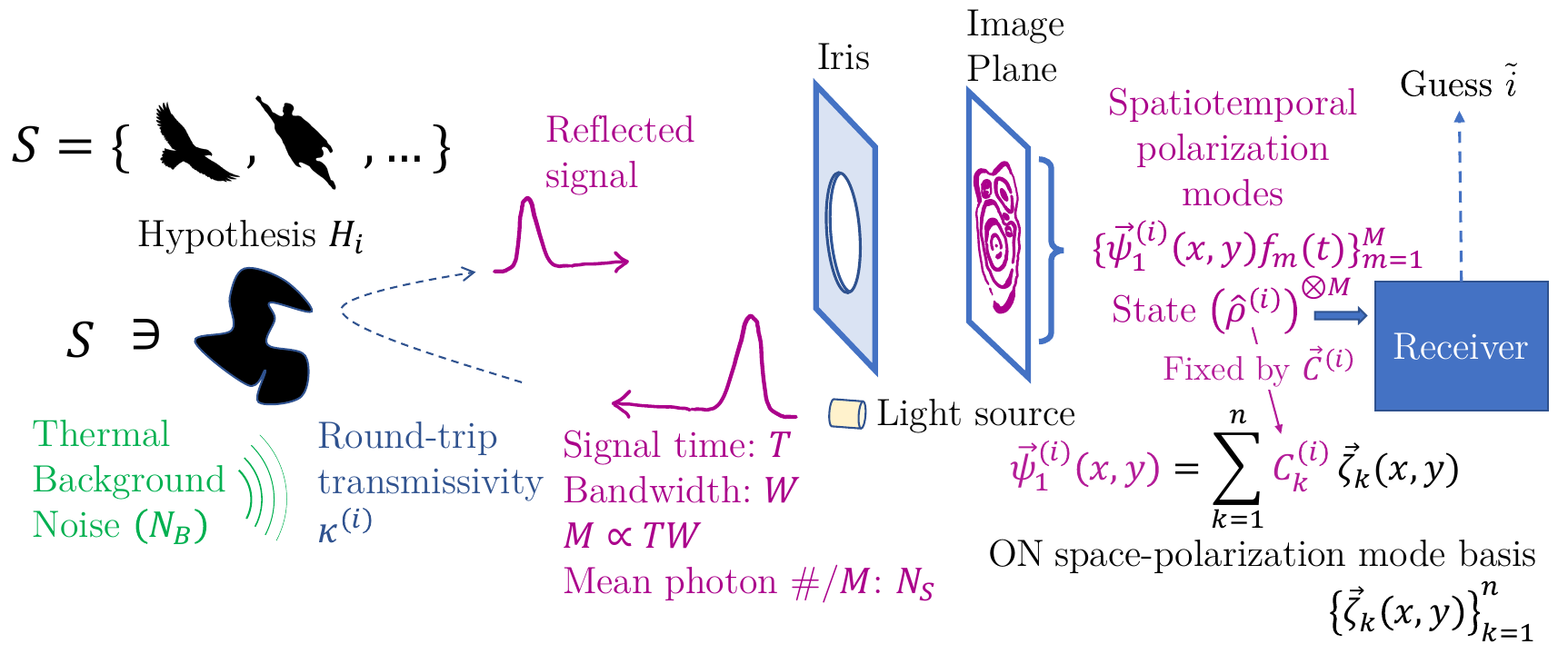}
\caption{Diagram illustrating the stand-off target discrimination task for a set of known targets $S$. The light emitted from the source is a signal of duration $T$ and bandwidth $W$, supporting $M<TW$ quasi-monochromatic frequency modes, which are initiated in a product of $M$ identical states of mean photon number $N_S$, leading to a product of $M$ identical states exciting $M$ orthonormal target-dependent spatiotemporal-polarization modes $\{\vec{\psi}_1^{(i)}(x,y)f_m(t)\}_{m=1}^M$ defined on the image plane. The targets are placed in a thermal background characterized by an approximately fixed Planck-law-governed mean photon number per mode $N_B$ over the spectral profile of the signal, and target $i$ is known to induce a round-trip transmissivity $\kappa^{(i)}$. The target dependence of the image plane modes is reflected in the spatial component $\psi^{(i)}_1$, which decomposes into a fixed orthonomal spatial mode set $\{\zeta_k(x,y)\}_{k=1}^n$ via the coefficient vector $\ora{C}^{(i)}.$ The identical states comprising the target return product state, when represented in the $n$-mode $\zeta$- basis, are denoted $\hat{\rho}^{(i)},$ with a one-to-one correspondence with $\ora{C}^{(i)}.$ The receiver performs a collective measurement on the product $\left(\hat{\rho}^{(i)}\right)^{\otimes M},$ the outcome on which a guess $\tilde{i}$ of the true hypothesis $H_i\in S$ is based.}
\label{fig:setup}
\end{figure}

    In the general version of the stand-off discrimination problem, which is depicted in Fig. \ref{fig:setup}, the goal is to successfully identify one of arbitrarily many and arbitrarily shaped non-radiating, matte, and reflective linear material targets drawn from a set $S$ by shining an electromagnetic (EM) radiation signal from a source of bandwidth $W$ and duration $T$ onto the target. The reflected EM field is collected through an iris, behind which there is an image plane where the light is detected by an optimal receiver. The targets are placed in a uniform thermal background, which contributes on average $N_B$ photons to the EM modes of the image plane in the spectral range of the signal. Let $\kappa^{(i)}$ be the probability that a photon emitted from the source makes it to the image plane after reflecting off of target $i$. Each target takes up space in the receiver's field of view, obscuring part of the thermal background. Since the targets are matte, reflective, and non-radiating, the probability that target $i$ blocks or redirects an environmental thermal photon exciting its target-return mode is proportional to $\kappa^{(i)},$ casting a thermal shadow even without active illumination. For simplicity, we take this probability to be equal to $\kappa^{(i)}.$ We assume the targets appear spectrally uniform, so that $\kappa^{(i)}$ is independent of the signal mode's frequency, depending only on the target's spatial configuration. 
    Then the channel corresponding to the journey undergone by a signal mode from source to target $i$ and back to the image plane can be described by a single beam splitter of transmissivity $\kappa^{(i)},$ serving also to couple in thermal light from an environment mode of mean photon number $N_B.$ 
    
    Reflective, for a target composed of a linear material, implies that the frequency range of width $W$ of the source is far off-resonance, so that the real index of refraction is mostly constant across the band, and hence the polarization of the reflected light is also frequency-independent.

The source signal is an EM field with a frequency support in range $\omega\in(\nu_0-W/2,\nu_0+W/2).$ It is modified by abruptly being turned on at a time $t_0,$ and abruptly turned off at time $t_0+T.$ In the limit of $T\rightarrow\infty,$ such a signal is known, per\cite{Land62}, to contain at most $\lfloor WT\rfloor$ independent degrees of freedom, saturated by the special basis of prolate spheroidal wave functions. In this case, we choose the signal to consist of the maximum number $M\leq\lfloor WT\rfloor$ of orthogonalized narrow-frequency pulses that fit into the time-bandwidth product of the signal. By choosing such quasi-monochromatic modes, we ensure that a signal excited in a tensor product of states across the $M$ frequency modes results in a product of $M$ states on the image plane on which the receiver acts. We label these frequency modes $\{f_m\}_{m=1}^M,$ and denote the space-polarization component of the source signal by $\vec{\psi}_0(x,y),$ and the space polarization component of the signal reflected from target $i$ by $\vec{\psi}_1^{(i)}(x,y).$ We take all modes to have unit $\mathcal{L}^2$-norm.

We can now describe the difference between the two types of illumination whose performance we seek to compare, namely GQI and CI. In CI, the signal modes $\{\vec{\psi}_0(x,y)f_m(t)\}_m$ are excited in an M-fold tensor product of coherent states, each of mean photon number $N_S.$ In GQI, these same signal modes are paired with a complementary set $\{\vec{\zeta}_I(x,y)f_m(t)\}_m$ of idler modes that are retained at the receiver, and the mode pairs $\{(\vec{\psi}_0(x,y)f_m(t),(\vec{\zeta}_I(x,y)f_m(t))\}_{m=1}^M,$ with $\vec{\zeta}_I$ spatially separated from $\vec{\psi}_0$ (and hence orthogonal to $\vec{\psi}_0$), are excited in an $M$-fold tensor product of two-mode-squeezed-vacuum (TMSV) with mean photon number per mode $N_S$. Both receivers employ mode sorting of the image plane light into a fixed, target-independent space-polarization mode basis $\{\vec{\zeta}_j(x,y)\}_{j=1}^\infty.$ For target $i,$ the state exciting the fixed sorted modes in CI is denoted $\hat{\rho}_C^{(i)},$ and in GQI, is denoted $\hat{\rho}_Q^{(i)}.$

\begin{figure}[ht]\centering
    \begin{subfigure}{.49\linewidth}
        \includegraphics[width=\linewidth]{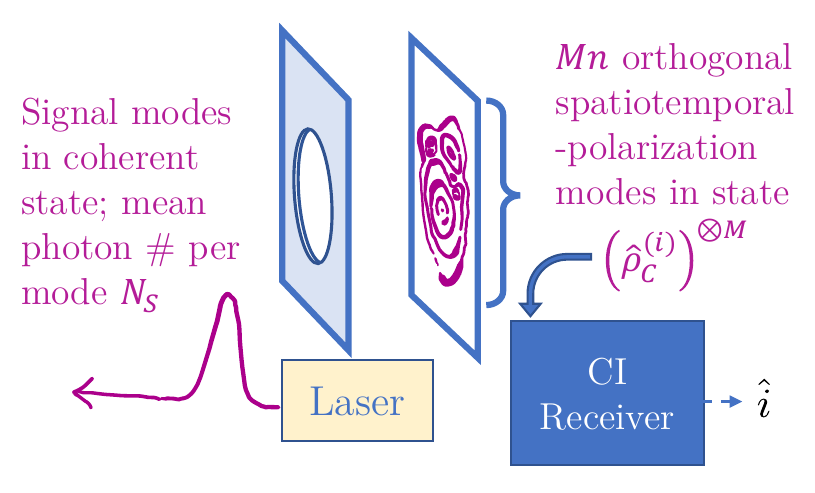}
        \caption{Classical illumination (CI)}
        \label{fig:CI_diagram}
    \end{subfigure}
    \begin{subfigure}{.49\linewidth}
        \includegraphics[width=\linewidth]{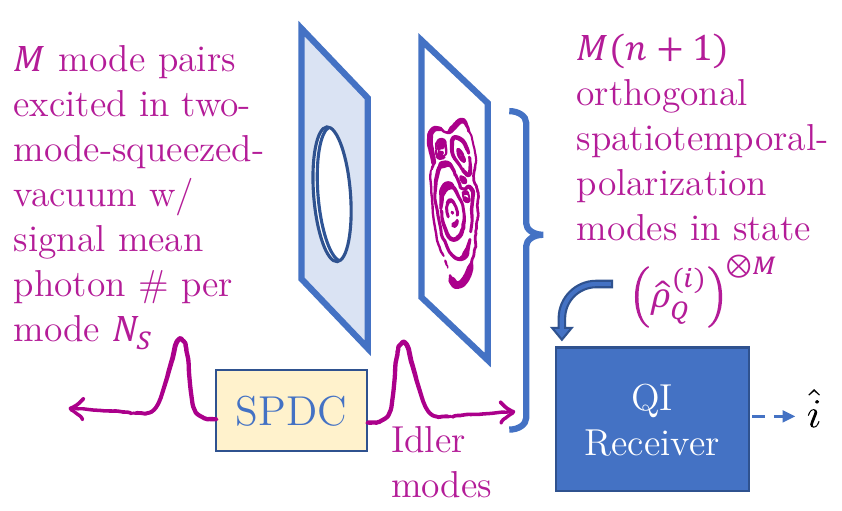}
        \caption{Quantum Illumination (QI)}
        \label{fig:QI_diagram}
    \end{subfigure}
    \caption{Comparison of the form of the target return light for CI and QI. For each frequency mode $m,$ the corresponding signal mode at the source is excited in (\subref{fig:CI_diagram}) a coherent state with real displacement $\sqrt{N_S},$ resulting in the $n$ receiver modes $\{\vec{\zeta}_k(x,y)f_m(t)\}_{k=1}^n$ being excited in the state $\hat{\rho}^{(i)}_C,$ and (\subref{fig:QI_diagram}) a signal-idler mode pair emitted from a SPDC source, excited in a two-mode-squeezed-vacuum of signal mean photon number $N_S$, resulting in the $n+1$ orthonormal receiver modes $\{\vec{\zeta}_k(x,y)f_m(t)\}_{k=1}^n\cup\{\vec{\zeta}_I(x,y)f_m(t)\}$ being excited in the state $\hat{\rho}^{(i)}_Q.$}\label{fig:setupcomparison}
\end{figure}

Assuming 1) the targets are sufficiently far from the source, and 2)
the iris opening is much larger than the distance from the iris to the image plane, $\vec{\psi}_1^{(i)}$ can be taken approximately parallel to the image plane, so the space-polarization basis elements can be chosen to be separable as a product $\vec{\zeta}_j(x,y)=\zeta_j(x,y)\hat{\epsilon}$ where $\hat{\epsilon}\in\{\hat{x},\hat{y}\},$ and the spatial Fourier transforms $\{\tilde{\zeta}_j(k_x,k_y)\}_j$ of $\{\zeta(x,y)_j\}_j$ form any orthonormal basis of $\mathcal{L}^2(\mathcal{R})$ where $\mathcal{R}$ is a disk of radius $k_{\text{max}}\ll \nu_0/c.$ So the target return light for CI is excited in a tensor product of $M$ copies of an $n$-mode state $\hat{\rho}_C^{(i)},$ for a total of $Mn$ spatiotemporal-polarization modes, and for GQI, the target return light, together with the light retained in the idler mode, is excited in $M$ copies of a $(n+1)$-mode state $\hat{\rho}_Q$, for a total of $M(n+1)$ spatiotemporal-polarization modes, as compared in Figs.~\ref{fig:setupcomparison}.
We can take $n$ to be as large as needed to span the set of spatial modes excited by the target return light under all hypotheses. We arrange the fixed space-polarization modes of the receiver in a sequence $\left(\vec{\zeta}_j\right)_j$ consisting of the $\zeta_j$ paired with alternating horizontal and vertical polarization modes: $$\vec{\zeta}_j(x,y):=\begin{cases}\hat{x}\zeta_{j/2}(x,y),&\text{$j$ even}\\\hat{y}\zeta_{\lceil j/2\rceil}(x,y),&\text{$j$ odd}\end{cases}$$ for all $j\in\{1,\dots,n\}.$ The image plane space-temporal modes cast by target $i$ can therefore be expanded as:
\begin{equation}
\vec{\psi}^{(i)}_{1}(x,y)=\sum_{j=1}^nC^{(i)}_j\vec{\zeta}_j(x,y),
\label{eq:objmodes}
\end{equation}
where the expansion coefficients $C^{(i)}_j\in\C$ for all $i,j$ have unit norm, i.e. $\sum_{j=1}^n|C^{(i)}_j|^2=1$, $\forall i$. Since the fixed spatio-polarization modes of the receiver form an orthonormal basis, the expansion coefficients $C^{(i)}_j$ can be evaluated as
\begin{multline}
C^{(i)}_j=\langle\vec{\psi}^{(i)}_1,\vec{\zeta}_j\rangle=\int_{\R^2}\vec{\psi}^{(i)}_1(x,y)\cdot\overline{\vec{\zeta}_j(x,y)}dxdy,
\label{eq:expansioncoeffs}
\end{multline}
where the bar denotes complex conjugation.

Let $\ora{C}^{(i)}$ denote the vector $\{C^{(i)}_1,C^{(i)}_2,\cdots,C^{(i)}_n\}.$ Note that $\ora{C}^{(i)}$ contains all information about target $i$, since $n$ is large enough such that $\psi^{(i)}_1$ is in $\text{Sp}\left(\{\zeta_j\}_{j=1}^n\right).$ Thus, the Chernoff exponents $\xi_C$ and $\xi_Q$ describing the decay rate of the minimum error probability using CI and GQI respectively, are purely a function of the set $\{\ora{C}^{(i)}\}_{i=1}^{|S|}.$

Since it was shown in \cite{Li16} that the Chernoff exponent for hypothesis testing with $|S|>2$ is equal to the Chernoff exponent for hypothesis testing between the most difficult to distinguish pair in $S,$ we let $|S|=2$ without loss of generality.

\section{State of Light in the Sorted Modes}\label{sec:state}
We start by setting the notation for the case of quantum illumination. Let $j,k\in\{1,2,\dots,2(n+1)\}$ and $i\in\{1,2\}$, and fix $m\in\{1,\dots,M\}.$ With the convention that $[\hat{a}_j,\hat{a}_k^\dagger]=\delta_{jk},$ for the annihilation operators associated with the spatio-temporal polarization modes $\{\vec{\zeta}_jf_m\}_j$, we define the position and momentum quadratures as $\hat{q}_j=\frac{1}{2}(\hat{a}_j+\hat{a}_j^\dagger)$ and $\hat{p}_j=\frac{1}{2i}(\hat{a}_j-\hat{a}_j^\dagger)$ for $j\in\{1,2,\dots,n\}\cup\{I\}$. We further define the phase-space coordinate vector as
$$\vec{\hat{r}}:=(\hat{q}_1\hat{q}_2,\dots,\hat{q}_n,\hat{q}_I,\hat{p}_1,\hat{p}_2,\dots,\hat{p}_n,\hat{p}_I),$$
so that $[\vec{\hat{r}},\vec{\hat{r}}]=\frac{i}{2}\Omega_{n+1}$ where $\Omega_n:=\begin{spmatrix}0&1\\-1&0\end{spmatrix}\otimes \mathbf{I}_n$. The mean vector components and Wigner covariance matrix elements for the states of the two objects is then given by
\begin{align}
(\mu^{(i)}_Q)_k=&\langle\hat{r}_k\rangle_i\label{eq:meanelts}\\
(V^{(i)}_Q)_{jk}=&\frac{1}{2}\langle\{\hat{r}_j,\hat{r}_k\}\rangle_i-\langle\hat{r}_j\rangle_i\langle\hat{r}_k\rangle_i.\label{eq:covelts1}
\end{align}
where $j,k\in\{1,2,\dots,2(n+1)\}$. The subscripted mean $\langle\cdot\rangle_i$ refers to taking the mean over the state in the sorted and idler modes for target $i$ - we denote this state by $\hat{\rho}_Q^{(i)}.$

The same conventions are followed for the case of classical illumination, with the only difference being the lack of the idler mode quadratures, so that $j$ and $k$ only runs from $\{1,2,\dots,2n\}$ in Eqs.~\eqref{eq:covelts1} and~\eqref{eq:meanelts}. In this case, we denote the mean vector, covariance matrix, and state of the light in the sorted modes for the $i^\text{th}$ object as $\vec{\mu}^{(i)}_C, V^{(i)}_C,$ and $\hat{\rho}^{(i)}_C$ respectively.

For arbitrary $n$, the mean vector and covariance matrices corresponding to quantum and classical illumination can be written compactly as
\begin{align}
    2V_C^{(i)}=&\left(N_B+\frac{1}{2}\right)\mb{I}_{2n}-\kappa^{(i)} N_B\;\mb{I}_{2}\otimes \ora{C}^{(i)}(\ora{C}^{(i)})^{\rm{T}}\label{eq:Vccompactfull}\\
    \mu_C^{(i)}=&\sqrt{\kappa^{(i)} N_S}(\ora{C}^{(i)},\vec{0}_n)\label{eq:mucfull}\\
    2V_Q^{(i)}=&\frac{1}{2}\mb{I}_{2(n+1)}+\overset{D}{\overbrace{\text{diag}(\underset{n\text{ many}}{\underbrace{N_B,\dots}},N_S,\underset{n\text{ many}}{\underbrace{N_B,\dots}},N_S)}}\nonumber\\
    -&\kappa^{(i)} (N_B-N_S)\overset{A^{(i)}}{\overbrace{\;\mb{I}_{2}\otimes \left((\ora{C}^{(i)},0)(\ora{C}^{(i)},0)^{\rm{T}}\right)}}\nonumber\\ +\mb{C_q}&\underset{B^{(i)}}{\underbrace{\begin{spmatrix}\vec{\delta}_{n+1}(\ora{C}^{(i)},\vec{0}_{n+2})^{\rm{T}}+(\ora{C}^{(i)},\vec{0}_{n+2})\vec{\delta}_{n+1}^{\rm{T}}\\-\vec{\delta}_{2(n+1)}(\vec{0}_{n+1},\ora{C}^{(i)},0)^{\rm{T}}-(\vec{0}_{n+1},\ora{C}^{(i)},0)\vec{\delta}_{2(n+1)}^{\rm{T}}\end{spmatrix}}}\label{eq:Vqcompactfull}\\
    \mu^{(i)}_Q=&\vec{0}_{2(n+1)}\label{eq:muqfull}
\end{align}
where $\mb{I}_k$ for $k\in\mbb{N}$ denotes the $k\times k$ identity matrix; $\text{\text{diag}(\dots)}$ denotes a diagonal matrix whose diagonal entries are given (in order from top left to bottom right) by the arguments; $\vec{0}_n$ is the vector of $n$ zeros; $\vec{\delta}_k$ is defined as a $2(n+1)$-dimensional vector with all entries zero except for the $k^\text{th}$ entry, which is 1; the notation $(\vec{v},\vec{u})$ for respectively $n$ and $m$-dimensional vectors $\vec{v}$ and $\vec{u}$ represents the vector $(v_1,\dots,v_n,u_1,\dots,u_m);$ all vectors are simultaneously treated as column matrices, and $\otimes$ denotes the Kronecker product of matrices. The groupings $D, A^{(i)}$ and $B^{(i)}$ in Eq.~\eqref{eq:Vqcompactfull} are  for ease of perturbative analysis going forward.

We wish to examine the asymptotic behavior of the Chernoff exponents when $N_S\rightarrow 0$ in the regime where $\kappa^{(i)}\ll N_S\ll 1\ll N_B.$ Note that in this regime, the state of light in the sorted modes is classical and admits a $P$-function representation. This can be seen by verifying that $V_Q^{(i)}-\mb{I}/4$ and $V_C^{(i)}-\mb{I}/4$ are positive definite, and hence the functions $f_Q:\R^{2(n+1)}\rightarrow\R$ and $f_C:\R^{2n}\rightarrow\R$ defined by $f_{Q}(\vec{r})=e^{-\vec{r}^{\rm T}(V_{Q}^{(i)}-\mb{I}/4)^{-1}\vec{r}}$ and $f_C(\vec{r})=e^{-\vec{r}^{\rm T}(V_{C}^{(i)}-\mb{I}/4)^{-1}\vec{r}}$ are normalizable. The states of the sorted modes corresponding to the $i^\text{th}$ target can then be written as
\begin{align}
    \hat{\rho}_C^{(i)}=&\int_{\R^{2n}}P_C^{(i)}\left(\vec{r}_C\right)\,\hat{\rho}_{\text{coh}}(\vec{x}_C,\vec{p}_C)\,\prod_{k=1}^ndx_kdp_k\label{eq:rhoc}\\
    \hat{\rho}_Q^{(i)}=&\int_{\R^{2(n+1)}}P_Q^{(i)}\left(\vec{r}_Q\right)\,\hat{\rho}_{\text{coh}}(\vec{x}_Q,\vec{p}_Q)\,dx_Idp_I\prod_{k=1}^ndx_kdp_k\label{eq:rhoq}.
\end{align}

The $P$-functions appearing in the integrands of Eq.~\eqref{eq:rhoc} and~\eqref{eq:rhoq} are given by
\begin{align}
    P_C^{(i)}\left(\vec{r}_C\right)=&\frac{e^{-\frac{1}{2}(\vec{r}_C-\vec{\mu}^{(i)}_C)^{\rm T}\left(V_C^{(i)}-\mb{I}/4\right)^{-1}(\vec{r}_C-\vec{\mu}_C^{(i)})}}{(2\pi)^n\sqrt{\det(V_C^{(i)}-\mb{I}/4)}}\label{eq:Pc}\\
    P_Q^{(i)}\left(\vec{r}_Q\right)=&\frac{e^{-\frac{1}{2}(\vec{r}_Q-\vec{\mu}^{(i)}_C)^{\rm T}\left(V_Q^{(i)}-\mb{I}/4\right)^{-1}(\vec{r}_Q-\vec{\mu}_C^{(i)})}}{(2\pi)^{n+1}\sqrt{\det(V_Q^{(i)}-\mb{I}/4)}}\label{eq:Pq},
\end{align}
where $\vec{r}_{C/Q}=\left(\vec{x}_{C/Q},\vec{p}_{C/Q}\right),$ with $\vec{x}_C,\vec{p}_C\in\R^n$ and $\vec{x}_Q,\vec{p}_Q\in\R^{n+1}$ such that the $(n+1)^\text{th}$ component of $\vec{x}_Q$ and $\vec{p}_Q$ correspond to the idler mode, and are labeled $x_I$ and $p_I$ respectively. $\hat{\rho}_{\text{coh}}(\vec{x}_{C/Q},\vec{p}_{C/Q})$ in Eqs.~\eqref{eq:rhoc} and~\eqref{eq:rhoq} represents the density matrix of a product of coherent states $(|x_k+\sqrt{-1}p_k\rangle)_k$ where $x_k+\sqrt{-1}p_k$ is the complex amplitude of a coherent state in the $k^\text{th}$ mode.

\section{Perturbative analysis of the Chernoff exponents}\label{sec:perturbation}
In the high loss and low signal brightness regime $\kappa^{(i)}\ll N_S\ll N_B$, the Chernoff exponents for classical and quantum illumination governing the error probabilities of distinguishing target 1 from target 2 become close to zero, since the mean vectors and covariance matrices given by Eqs.~\eqref{eq:Vccompactfull}-\eqref{eq:muqfull} show very weak dependence on the target index $i$. In this section, we aim to obtain asymptotic expressions for $\xi_Q$ and $\xi_C$ for the quantum and classical Chernoff exponents respectively in terms of $\kappa^{(i)}, N_S$ and $N_B$ such that $\xi_Q=\xi_C=0$ when $\kappa^{(i)}=N_S=0$ for all $i\in\{1,2\}$. 

To do this, we choose a restricted direction of approach to the limit $\kappa^{(i)}, N_S\rightarrow 0$ and $N_B\rightarrow\infty,$ namely\footnote{We use asymptotic notation for positive arguments of functions approaching zero rather than positive infinity, so $f\in\mathcal{O}(g)$ if  $\exists c\in\R^{+}$ such that $\exists x_0\in\R^{+}$  for which $f(x)<cg(x)$ for all positive $x\leq x_0,$ and $f\in o(g)$ if $\forall c\in\R^{+},$ $\exists x_0\in\R^{+}$ such that $f(x)<cg(x)$ for all positive $x\leq x_0.$}  
\begin{equation}
\begin{gathered}
    \kappa^{(i)}\in o(N_S/N_B)\\
    N_S\in\mathcal{O}(\exp(-N_B))\\
    1\in o(N_B),\\
\end{gathered}\label{eq:parameterscalingminreqs}\end{equation}
where $\kappa^{(i)}, N_S,$ and $N_B$ are taken to be functions of a path parameter $t,$
and start with an expansion of the states given by Eqs.~\eqref{eq:rhoc} and~\eqref{eq:rhoq} in the form
\begin{align}    \hat{\rho}_C^{(i)}\approx\hat{\rho}_{C}^{(0)}+\hat{\nu}_C^{(i)}\label{eq:rhocperturbedform}\\    \hat{\rho}_Q^{(i)}\approx\hat{\rho}_{Q}^{(0)}+\hat{\nu}_Q^{(i)},\label{eq:rhoqperturbedform}
\end{align}
such that $\rho^{(0)}_{C/Q}$ is a product of thermal states and the target dependence is only on the traceless operators $\hat{\nu}_{C/Q}^{(i)},$ which are the lowest-order terms perturbing $\hat{\rho}_C^{(i)}$ and $\hat{\rho}_Q^{(i)},$ and which vanish in the limit of $\kappa^{(i)},\,N_S\rightarrow 0.$ The expansions \eqref{eq:rhocperturbedform} and \eqref{eq:rhoqperturbedform} can be used in a perturbative formula for the Chernoff exponents from \cite{Grace22}:
\begin{equation}
    \xi_{C/Q}\approx\frac{1}{2}\text{Tr}\left[L_{\sqrt{x}}\left(\hat{\rho}^{(0)}_{C/Q},\hat{\nu}^{(1)}_{C/Q}-\hat{\nu}^{(2)}_{C/Q}\right)^2\right]
    \label{eq:michaelschernoff}
\end{equation}
where $L_{\sqrt{x}}(A,B)$ denotes the operator-valued Frechet derivative of the function $\sqrt{x}$ for a perturbation $B$ around $A.$
$\hat{\rho}^{(0)},\,\hat{\nu}^{(1)}_{C/Q},$ and $\hat{\nu}^{(2)}_{C/Q}$ in Eq.~\eqref{eq:michaelschernoff} can be obtained by expanding the $P$-functions given by Eqs.~\eqref{eq:Pc} and~\eqref{eq:Pq} in the high loss and low signal brightness regime. It can be shown, by recursive use of the determinant property $\det(\mathbf{I}+\vec{v}\vec{u}^{\tr})=1+\vec{u}^\tr\vec{v}$ for real vectors $\vec{v}$ and $\vec{u},$ that that
\begin{multline}
    \det\left(V_Q-\mb{I}/4\right)^{-1/2}\\=2^{n+1}\det(D)^{-1/2}\left(1+\kappa^{(i)} +\mathcal{O}\left(\frac{\kappa^{(i)} N_S}{N_B}\right)\right),\label{eq:detVqminusIinv}
\end{multline}

and by a Neumann series expansion,
\begin{multline}
    \frac{1}{2}(\vec{r}_Q-\vec{\mu}_Q^{(i)})^{\rm T}\left(V_Q-\mb{I}/4\right)^{-1}(\vec{r}_Q-\vec{\mu}_Q^{(i)})\\=\frac{\|\vec{r}_C\|^2}{N_B}+\frac{x_I^2+p_I^2}{N_S}-2\frac{\sqrt{\kappa^{(i)}/N_S}}{N_B}(x_I\vec{x}_C-p_I\vec{p}_C)\cdot\ora{C}^{(i)}\\+\mathcal{O}\left(\frac{\kappa^{(i)}}{N_SN_B}\right).\label{eq:Pfuncexponentexpanded}
\end{multline}
Then, using the Taylor series expansion $e^{x+\varepsilon}=e^x(1+\varepsilon+\mathcal{O}(\varepsilon^2)),$
\begin{multline}
    P_Q^{(i)}(\vec{r}_Q)=\frac{e^{-\frac{|\vec{x}_C|^2+|\vec{p}_C|^2}{N_B}-\frac{x_I^2+p_I^2}{N_S}}}{\pi^{n+1}\sqrt{\det{D}}}\times\\\left(1-\sqrt{\frac{\kappa^{(i)}}{N_S}}\frac{2}{N_B}\ora{C}^{(i)}\cdot(x_I\vec{x}_C-p_I\vec{p}_C)+\mathcal{O}\left(\frac{\kappa^{(i)}}{N_SN_B}\right)\right).\label{eq:Pqexpansion}
\end{multline}
Similarly, the $P$-function for the classical illumination task can be expanded starting from Eqs.\eqref{eq:Vccompactfull} and~\eqref{eq:mucfull}, and applying the Neumann series expansion to obtain
\begin{multline}
    \left(V_C-\mb{I}/4\right)^{-1}\\=\frac{2}{N_B}\left(\mb{I}_{2n}+\kappa^{(i)}\mb{I}_2\otimes\ora{C}^{(i)}(\ora{C}^{(i)})^{\rm T}+\mathcal{O}\left(\left(\kappa^{(i)}\right)^2\right)R\right),\label{eq:VcminusIinvexpanded}
\end{multline}
where $R$ is a residual matrix in which each entry is $-1, 0,$ or $1.$ Then
\begin{multline}
    \frac{1}{2}(\vec{r}_C-\vec{\mu}_C^{(i)})^{\rm T}\left(V_C-\mb{I}/4\right)^{-1}(\vec{r}_C-\vec{\mu}_C^{(i)})\\=\frac{\|\vec{r}\|_C^2}{N_B}-2\frac{\sqrt{\kappa^{(i)} N_S}}{N_B}\vec{x}_C\cdot\ora{C}^{(i)}+\mathcal{O}\left(\kappa^{(i)}/N_B\right),\label{eq:Pcexponentexpanded}
\end{multline}
and the $P$-function normalization factor can be expanded as
\begin{equation}
    \det\left(V_C-\mb{I}/4\right)^{-1/2}=2^nN_B^{-n}\left(1+\mathcal{O}(\kappa^{(i)}/N_B)\right).\label{eq:detVcminusIinv}
\end{equation}
Once again using the Taylor series expansion $e^{x+\varepsilon}=e^x(1+\varepsilon+\mathcal{O}(\varepsilon^2)),$ the P-function of the received state of light for CI expands as
\begin{multline}
    P_C^{(i)}(\vec{r}_C)=\frac{e^{-\frac{\|\vec{r}_C\|^2}{N_B}}}{\pi^n N_B^n}\\\times\left(1-\sqrt{\kappa^{(i)} N_S}\frac{2}{N_B}\ora{C}^{(i)}\cdot\vec{x}_C+\mathcal{O}(\kappa^{(i)}/N_B)\right).\label{eq:Pcexpansion}
\end{multline}
Comparing the form given by Eq.~\eqref{eq:rhocperturbedform} with Eq.~\eqref{eq:Pcexpansion}, and the form given by Eq.~\eqref{eq:rhoqperturbedform} with Eq.~\eqref{eq:Pqexpansion}, we identify:
\begin{subequations}
\begin{align}
    \hat{\rho}_C^{(0)}=&\frac{1}{\pi^nN_B^n}\int_{\R^{2n}}e^{-\frac{\|\vec{r}_C\|^2}{N_B}}\hat{\rho}_{\text{coh}}(\vec{x}_C,\vec{p}_C)\prod_{k=1}^ndx_kdp_k\\
    \hat{\nu}_C^{(i)}=&\frac{-2}{\pi^nN_B^n}\int_{\R^{2n}}e^{-\frac{\|\vec{r}_C\|^2}{N_B}}\frac{\sqrt{\kappa^{(i)} N_S}}{N_B}\nonumber\\&\hspace{3.5em}\times\vec{x}_C\cdot\ora{C}^{(i)}\hat{\rho}_{\text{coh}}(\vec{x}_C,\vec{p}_C)\prod_{k=1}^ndx_kdp_k\\
    \hat{\rho}_Q^{(0)}=&\frac{1}{\pi^{n+1}\sqrt{|D|}}\int_{\R^{2(n+1)}}e^{-\frac{\|\vec{r}_C\|^2}{N_B}-\frac{x_I^2+p_I^2}{N_S}}\nonumber\\&\hspace{4em}\times\hat{\rho}_{\text{coh}}(\vec{x}_Q,\vec{p}_Q)dx_Idp_I\prod_{k=1}^ndx_kdp_k\\
    \hat{\nu}_Q^{(i)}=&\int_{\R^{2(n+1)}}\frac{e^{-\frac{\|\vec{r}_C\|^2}{N_B}-\frac{x_I^2+p_I^2}{N_S}}}{\pi^{n+1}\sqrt{|D|}}\left(-\sqrt{\frac{\kappa^{(i)}}{N_S}}\frac{2}{N_B}\right)\ora{C}^{(i)}\nonumber\\&\hspace{-2em}\cdot(x_I\vec{x}_C-p_I\vec{p}_C)\hat{\rho}_{\text{coh}}(\vec{x}_Q,\vec{p}_Q)dx_Idp_I\prod_{k=1}^ndx_kdp_k.
\end{align}\label{eq:alloperatorexpansionterms}
\end{subequations}
Asymptotic expansions of the quantum Chernoff exponents can now be evaluated using Eq.~\eqref{eq:michaelschernoff}. In the eigensystem of $\hat{\rho}_{C/Q}^{(0)},$ consisting of eigenvalues $\{\lambda_{\vec{m}}\}_{\vec{m}}$ (denoted collectively by vector $\vec{\lambda}$) corresponding to Fock product states $\{|\vec{m}\rangle\}_{\vec{m}},$ the first Frechet derivative is represented as
\begin{multline}
    L_{\sqrt{x}}\left(\hat{\rho}_{C/Q}^{(0)},\hat{\nu}_{C/Q}^{(1)}-\hat{\nu}_{C/Q}^{(2)}\right)=\\\sum_{\vec{m},\vec{m}'}\underset{\vec{m},\vec{m}'}{[\sqrt{x},\vec{\lambda}]}\langle\vec{m}|\hat{\nu}_{C/Q}^{(1)}-\hat{\nu}_{C/Q}^{(2)}|\vec{m}'\rangle|\vec{m}\rangle\langle\vec{m}'|,\label{eq:Frechet1}
\end{multline}
where $[\sqrt{x},\vec{\lambda}]$ denotes the first divided difference of $\sqrt{x}$ at $\hat{\rho}_{C/Q}^{(0)}$, given by
\begin{equation}
    \underset{\vec{m},\vec{m}'}{[\sqrt{x},\vec{\lambda}]}=\begin{cases}\frac{\sqrt{\lambda_{\vec{m}}}-\sqrt{\lambda_{\vec{m}'}}}{(\lambda_{\vec{m}}-\lambda_{\vec{m}'})},&\lambda_{\vec{m}}\neq\lambda_{\vec{m}'}\\\left.\frac{d}{dx}\sqrt{x}\right|_{x=\lambda_{\vec{m}}},&\text{otherwise.}\end{cases}\label{eq:fdd}
\end{equation}
Eq.~\eqref{eq:michaelschernoff}, evaluated by squaring, tracing and simplifying  Eq.~\eqref{eq:Frechet1}, gives
\begin{align}
    \xi_C\approx&\frac{ N_S}{4N_B}\left\|\sqrt{\kappa^{(2)}}\ora{C}^{(2)}-\sqrt{\kappa^{(1)}}\ora{C}^{(1)}\right\|^2.\label{eq:classicalchernofffinal}\\
    \text{Similarly, }\xi_Q\approx&\frac{ N_S}{N_B}\left\|\sqrt{\kappa^{(2)}}\ora{C}^{(2)}-\sqrt{\kappa^{(1)}}\ora{C}^{(1)}\right\|^2.
    \label{eq:quantumchernofffinal}
\end{align}

Eqs.~\eqref{eq:classicalchernofffinal}~and~\eqref{eq:quantumchernofffinal} prove that in the regime of $\kappa^{(i)}\in o(N_S/N_B),$ $N_S\in\mathcal{O}(\exp(-N_B)),$ and $1\in o(N_B),$
\begin{equation}
    \lim_{t\rightarrow 0}\frac{\xi_Q}{\xi_C}=4.\label{eq:ratioresult}
\end{equation}

The norms appearing Eq.~\eqref{eq:quantumchernofffinal} and~\eqref{eq:classicalchernofffinal} can be promoted to the general case of $\ora{C}^{(i)}\in\C^n$ by noting that a unitary transformation of the sorting coefficients $\ora{C}^{(i)}$ implies a target-independent unitary transformation of $\hat{\rho}_{C/Q}^{(i)},$ and that\cite{Grace22} the optimal exponent appearing in the general quantum chernoff exponent formula $\xi=\underset{s}{\sup}-\log \text{Tr}\left(\left(\hat{\rho}^{(1)}\right)^s\left(\hat{\rho}^{(2)}\right)^{(1-s)}\right)$ reduces to the Bhattacharya exponent when $\hat{\rho}^{(i)}$ for $i\in\{1,2\}$ is described by a small perturbation of a common ground state.

\section{Conclusion}\label{sec:conclusion}
We have shown that for the task of discriminating among a set of apriori known yet arbitrary reflective objects using an active-illumination radar at stand-off range, Gaussian quantum illumination---using an entangled spontaneous-parametric downconversion (SPDC) source---offers a factor-of-four advantage in the quantum Chernoff exponent over that of classical coherent-state illumination, in an appropriate limit of high return-path loss, low mean signal photon number per mode, and high mean thermal noise photon number per mode. To derive this result, we expressed a multi-spatial-mode span of the received light and the retained idler as a target-profile independent multi-mode thermal state perturbed by a traceless operator with a small norm, and that in the aforesaid regime the return-path channel is entanglement breaking, rendering the target-return plus idler joint state expressible using P functions. Our analysis reaffirms that a quantum illumination radar for use in large stand-off range has limited use due to the restrictive regime where there is a benefit over a classical alternative. Despite being entanglement breaking, that an entangled probe outperforms all classical probes, makes it an intriguing problem. The optimal receiver design is left open.

\ifCLASSOPTIONcaptionsoff
\fi
  \newpage

\end{document}